\documentclass[aps,prapplied,twocolumn,superscriptaddress]{revtex4-2}

\usepackage{graphicx} 
\usepackage{amsmath} 
\usepackage{mhchem} 
\usepackage[colorlinks,linkcolor=blue,citecolor=green,urlcolor=blue]{hyperref} 
\usepackage{color} 

\begin{document}


\title{Quantitative Assessment of Carrier Density by Cathodoluminescence. II. GaAs nanowires}


\author{Hung-Ling Chen}
\affiliation{Centre de Nanosciences et de Nanotechnologies (C2N), CNRS, Universit\'e Paris-Saclay, 91120 Palaiseau, France}

\author{Romaric De L\'epinau}
\affiliation{Centre de Nanosciences et de Nanotechnologies (C2N), CNRS, Universit\'e Paris-Saclay, 91120 Palaiseau, France}
\affiliation{Institut Photovolta\"ique d'Ile-de-France (IPVF), 91120 Palaiseau, France}

\author{Andrea Scaccabarozzi}
\author{Fabrice Oehler}
\author{Jean-Christophe Harmand}
\author{Andrea Cattoni}
\affiliation{Centre de Nanosciences et de Nanotechnologies (C2N), CNRS, Universit\'e Paris-Saclay, 91120 Palaiseau, France}

\author{St\'ephane Collin}
\affiliation{Centre de Nanosciences et de Nanotechnologies (C2N), CNRS, Universit\'e Paris-Saclay, 91120 Palaiseau, France}
\affiliation{Institut Photovolta\"ique d'Ile-de-France (IPVF), 91120 Palaiseau, France}
\email[]{stephane.collin@c2n.upsaclay.fr}


\date{\today}

\begin{abstract}
	
Precise control of doping in single nanowires (NWs) is essential for the development of NW-based devices. Here, we investigate a series of MBE-grown GaAs NWs with Be (p-type) and Si (n-type) doping using high-resolution cathodoluminescence (CL) mapping at low- and room-temperature. CL spectra are analyzed selectively in different regions of the NWs. Room-temperature luminescence is fitted with the generalized Planck law and an absorption model, and the bandgap and band tail width are extracted. For Be-doped GaAs NWs, the bandgap narrowing provides a quantitative determination of the hole concentration ranging from about $1\times 10^{18}$ to $2\times 10^{19}$~cm$^{-3}$, in good agreement with the targeted doping levels. High-resolution maps of the hole concentration demonstrate the homogeneous doping in the pure zinc-blende segment. For Si-doped GaAs NWs, the electron Fermi level and the full-width at half maximum of low-temperature CL spectra are used to assess the electron concentration to approximately $3\times 10^{17}$ to $6\times 10^{17}$~cm$^{-3}$. These findings confirm the difficulty to obtain highly-doped n-type GaAs NWs, maybe due to doping compensation. Notably, signatures of high concentration (5--9$\times 10^{18}$~cm$^{-3}$) at the very top of NWs are unveiled.

\end{abstract}


\maketitle


\section{Introduction}

Semiconductor nanostructures are widely investigated in fundamental research and for device applications. In particular, III-V nanowires (NWs) have attracted a growing interest for the development of optoelectronic devices such as laser~\cite{Hill:2007,Li:2015a,Eaton:2016}, light-emitting diodes~\cite{Svensson:2008,Tomioka:2010,Guan:2016}, photodetectors~\cite{Dai:2014,LaPierre:2017} and solar cells~\cite{Mariani:2013,Wallentin:2013,Aberg:2016,vanDam:2016}. For instance, the ability to grow defect-free III-V NWs on silicon (Si) allows for the direct growth of III-V/Si tandem solar cells~\cite{Yao:2015}. Doping is a key element for optimal operation of the devices like low-resistivity of the contacts~\cite{Huttenhofer:2018} or efficient carrier collection~\cite{Li:2015}. Owing to the small dimensions of NWs, the realization of radial or axial p-n junctions requires a very tight control of the doping conditions during the epitaxial growth. However, characterization of carrier densities at the sub-micrometer scale remains challenging.

Field effect measurements~\cite{Cui:2000,Dayeh:2006} and Hall effect measurements~\cite{Blomers:2012,Storm:2012} have been used to assess the transport properties of single NWs, but they require the fabrication of nanoscale electrical contacts, which is technically difficult. Multiple probes in an electron microscope have been used to examine the charge carrier depletion and doping profile along NWs~\cite{Nagelein:2018,Piazza:2018a,Saket2020}. Off-axis electron holography employs electron interference to measure the electrical potential in a thin specimen~\cite{Yazdi:2015}, and have been used to reveal doping inhomogeneities in GaAs NWs~\cite{Dastjerdi:2017,Hakkarainen:2019}. Time-of-flight secondary ion mass spectroscopy (SIMS)~\cite{Chia:2015} and atom probe tomography (APT)~\cite{Du:2013} can also be used to measure the density of impurity atoms, which can be much higher than the carrier concentration in case of electrically non-activated or compensated doping impurities. Alternatively, optical methods provide rapid and contactless analysis of samples without additional processing step. For instance, the collective oscillation of free carriers can be excited and measured in the terahertz domain, but the low spatial resolution limits this technique to the investigation of NW ensembles~\cite{Joyce:2016}. Raman spectroscopy allows to probe carrier densities through longitudinal optical (LO) phonon-plasmon interactions~\cite{Jeganathan:2009,Mohajerani:2016}, and can be combined with APT to infer the activation efficiency of dopant elements~\cite{Hakkarainen:2019}. Light scattering from local vibrational modes of Si atoms in a GaAs lattice may also indicate whether Si dopants occupy As sites (p-type doping)~\cite{Ketterer:2010,Ketterer:2012} or Ga sites (n-type doping)~\cite{Dimakis:2012}. Photoluminescence (PL) is widely used and the Burstein-Moss shift was observed in n-InP~\cite{Lindgren:2015}, n-GaAs~\cite{Arab:2016} and n-InAs~\cite{Sonner:2018} NWs. However, these optical techniques do not provide the spatial resolution required to study inhomogeneous NWs.

Here, we use cathodoluminescence (CL) mapping as a rapid, contactless method for the determination of carrier densities in single NWs. The high spatial resolution of this technique allows to distinguish different regions of a single NW. It is used to discard defective zones and to study variations of the carrier concentration in well-controlled homogeneous regions. In a previous work, we have demonstrated the quantitative analysis of electron concentration in Si-doped GaAs NWs grown randomly on a Si(111) substrate~\cite{Chen:2017}. Luminescence analysis method and its accuracy have been further validated on a series of p-type and n-type GaAs planar reference thin films~\cite{Chen:2019a}. Here, we extend this work to a series of Be-doped (core and shell) and Si-doped (shell) GaAs NWs grown on periodically patterned Si substrates with different doping level targets. We assess quantitatively the carrier densities of both p-type and n-type NWs, either by fitting room-temperature (RT) CL spectra with the generalized Planck law and an absorption model, or with the full-width at half maximum (FWHM) of low-temperature (LT) CL spectra. We discuss the agreement (p-doped) and disagreement (n-doped) with the target doping levels in different growth conditions, and the homogeneity of the carrier concentration. In GaAs:Si NWs, doping compensation limits the carrier concentration in the shell. Importantly, we unveil signatures of high electron concentrations in the tip segment.

\section{\label{sec:exp}Experimental details}

\begin{table*}
	\caption{\label{tab:sample}Description of doped GaAs NW samples. The growth temperature and the V/III BEP ratio are given. The NW length, diameter and shell thickness are obtained by examining SEM images (mean value $\pm$ standard deviation). The growth rate refers to the elongation of the NW length (core doping) or the expansion of the shell thickness (shell doping) per unit of time. The dopant flux was calibrated using GaAs reference thin films and is given with respect to the substrate normal. Theoretical doping concentrations are deduced from the dopant flux and the NW growth rate.}
	\begin{ruledtabular}
	\begin{tabular}{lcccccccc}
		sample & growth T. & V/III ratio & length & diameter & shell thick. & growth rate & dopant & th. doping conc. \\
		& $^\circ$C & & $\mu$m & nm & nm & $\mathring{A}$/s & atoms/(s$\cdotp$cm$^2$) & cm$^{-3}$ \\
		\hline
		core-1 (Be)  & 610 & 10  & 2.45$\pm$0.11 & 189$\pm$13 &            & 14.3$\pm$0.7    & $1.5\times 10^{11}$ & $(1.0\pm0.1)\:10^{18}$ \\ 
		core-2 (Be)  & 610 & 10  & 1.85$\pm$0.12 & 182$\pm$12 &            & 10.8$\pm$0.7    & $8.8\times 10^{11}$ & $(8.1\pm0.6)\:10^{18}$ \\ 
		shell-1 (Be) & 550 & 129 & 2.00$\pm$0.12 & 236$\pm$8  & 67$\pm$4   & 0.19$\pm$0.02 & $2.1\times 10^{10}$ & $(1.0\pm0.2)\:10^{18}$ \\ 
		shell-2 (Be) & 550 & 129 & 1.03$\pm$0.04 & 282$\pm$16 & 90$\pm$8   & 0.25$\pm$0.03 & $1.5\times 10^{11}$ & $(5.4\pm0.8)\:10^{18}$ \\ 
		shell-3 (Be) & 550 & 129 & 1.99$\pm$0.10 & 334$\pm$42 & 116$\pm$21 & 0.32$\pm$0.06 & $8.8\times 10^{11}$ & $(2.5\pm0.6)\:10^{19}$ \\
		\hline
		shell-4 (Si) & 450  & 153 & 2.06$\pm$0.10 & 292$\pm$14 & 95$\pm$7  & 0.26$\pm$0.03 & $4.6\times 10^{10}$ & $(3.9\pm0.4)\:10^{18}$ \\
		shell-5 (Si) & 475  & 117 & 2.15$\pm$0.06 & 285$\pm$22 & 91$\pm$11 & 0.19$\pm$0.03 & $5.5\times 10^{10}$ & $(6.4\pm1.2)\:10^{18}$ \\
	\end{tabular}
	\end{ruledtabular}
\end{table*}

NWs were grown on a patterned Si(111) substrate by molecular beam epitaxy (MBE). A \ce{SiO2} film (thickness $\sim$25~nm) was deposited by plasma-enhanced chemical vapor deposition (PECVD). It is used as a mask for the selective growth of NWs. Openings were fabricated using electron-beam lithography in a polymethyl methacrylate (PMMA) resist. They consist in a hexagonal array of holes (diameter $\sim$20~nm) with a pitch of 500~nm or 1~$\mu$m. The holes were transferred to the \ce{SiO2} mask by reactive-ion etching (RIE) with \ce{SF6}/\ce{CHF3} gases, followed by a wet chemical etching in dilute HF:\ce{H2O} (1:100) to remove the last nanometers of \ce{SiO2} before loading the substrate into the MBE chamber for degassing in ultra-high vacuum.

\begin{figure}
	\includegraphics[width=\columnwidth]{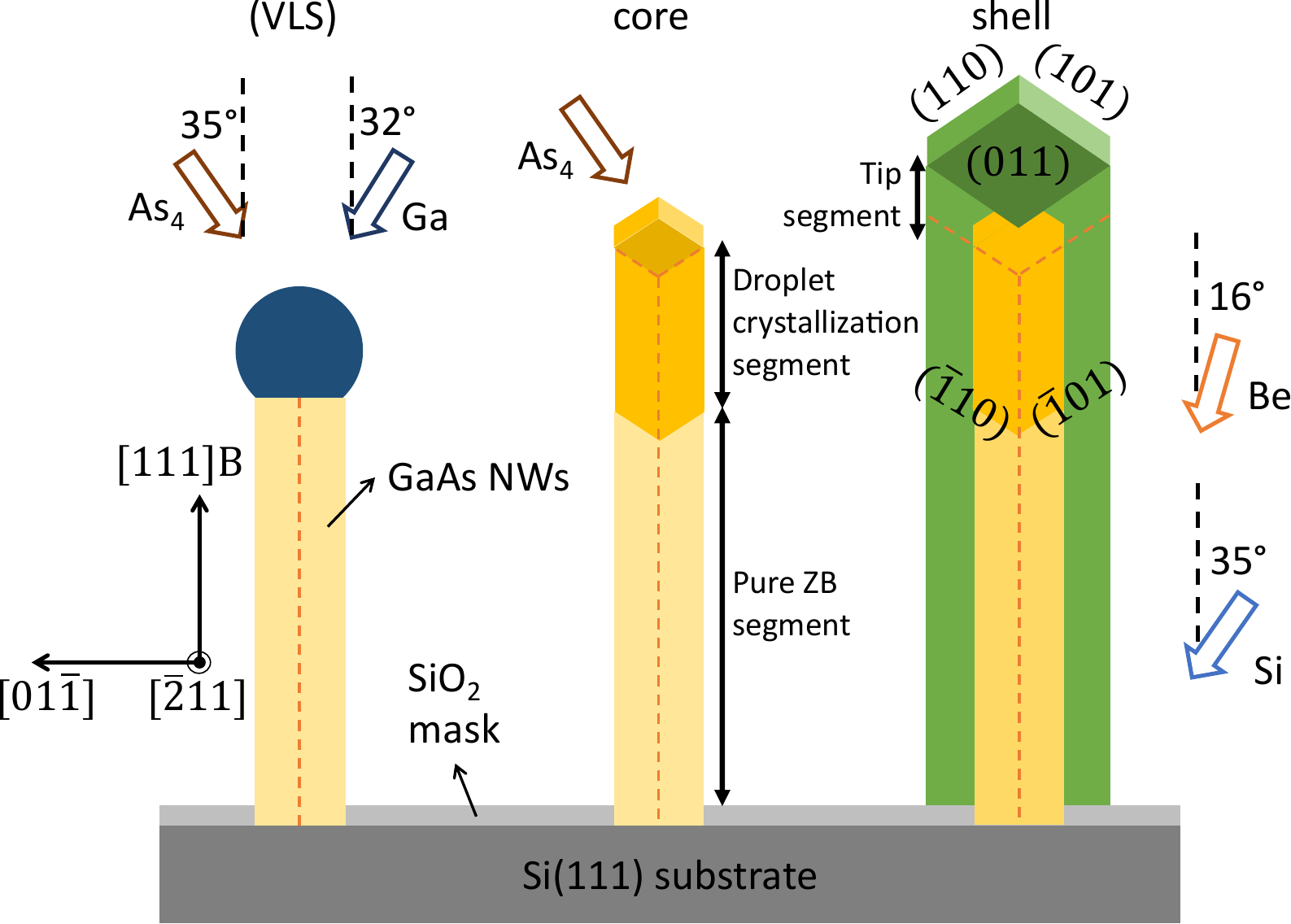}
	\caption{\label{fig:nw_growth}Overview of the NW growth sequences. NW cores are grown using the vapor-liquid-solid (VLS) method on a Si(111) substrate patterned with a \ce{SiO2} mask. Ga liquid droplets act as a catalyst of the MBE growth. They are crystallized at the end of the VLS growth. Subsequently, shell growth can be conducted (vapor-solid, VS growth) for surface passivation or to create a core-shell junction. The incident angles of molecular beams with respect to the substrate normal are indicated.}
\end{figure}

The MBE growth was performed in a Riber Compact 32 system. The NW growth sequences and the angles of molecular beams are depicted in Fig.~\ref{fig:nw_growth}. Ga was supplied using a standard effusion cell. It first results in localized liquid Ga droplets above the mask openings. Arsenic was then supplied as \ce{As4} by a valved source and resulted in the NW growth via the self-catalyzed vapor-liquid-solid (VLS) method. The VLS growth was conducted at 610$^\circ$C. After the VLS growth, Ga droplets were crystallized by closing the Ga shutter and increasing the As flux to roughly twice the previous value at a substrate temperature of about 500$^\circ$C (except for sample shell-4, see section V). This growth sequence yields a first, long zinc-blende (ZB) segment, named \emph{pure ZB segment}, and a second segment with mixed wurtzite (WZ) and ZB crystal structures (see transmission electron microscope images in the appendix, and the schematic in Fig.~\ref{fig:nw_growth}). This WZ/ZB mixed segment is attributed to the droplet crystallization~\cite{Priante2013,Kim2012a}, it represents approximately 30\% of the total length and it is named \emph{droplet crystallization segment}.

For two samples (core-1 and core-2), Be dopants were provided during the VLS growth and an AlGaAs passivation shell (thickness $\sim$13~nm) was grown after the droplet crystallization. For the others (shell-1 to shell-5), the NW cores were non intentionally doped, and doped shells were grown upon the sidewall facets of the \{110\} family in the conventional vapor-solid (VS) mode with Be (p-type) or Si (n-type) dopants. Detailed growth parameters are given in Table~\ref{tab:sample}, including the growth temperature and the V-to-III beam equivalent pressure (BEP) ratio used for growing the doped core or shell. The very top of shell-doped NWs, corresponding to the shell overgrowth on top of the core, is named \emph{tip segment}.

The total length and diameter of NWs are measured from scanning electron microscope (SEM) images and averaged over several NWs of the same sample. Despite the hexagonal shape of NWs, their diameter is referred to as the diameter of an equivalent cylinder with a cross-section of equal surface. The shell thickness is obtained by subtracting the typical diameter of NW core only (105$\pm$10~nm) to the total diameter. The flux of Be- and Si-dopants was previously calibrated on GaAs(001) thin films. For core-1 and core-2, the theoretical doping concentration is calculated as the dopant flux divided by the NW growth rate (the growth rate is estimated as 70\% of the total length divided by the growth time, which excludes the droplet crystallization). For shell-1 to shell-5, the theoretical doping concentration is obtained similarly but taking into account the projection of the dopant flux to the vertical sidewalls and the rotation symmetry of the shell growth.

After growth, NWs were dispersed on a Si substrate. CL measurements were performed in an Attolight Chronos quantitative cathodoluminescence microscope. The e-beam acceleration voltage is 6~kV and the impinging current is 0.3--0.7~nA. The luminescence is collected by an achromatic reflective objective (NA: 0.72) and analyzed with a Horiba iHR320 monochromator (grating: 150~grooves/mm) and an Andor Newton Si CCD camera (1024$\times$256 pixels, pixel width 26~$\mu$m, spectral dispersion: 0.53~nm per pixel). Luminescence spectra were corrected for the diffraction efficiency of the grating and the sensitivity of the CCD camera.
The spatial resolution of CL maps is limited by electron beam diameter (below 10 nm) and the interaction volume in GaAs, estimated to less than 40 nm in diameter and 70 nm in depth. Diffusion of charge carriers contributes to decrease the actual spatial resolution, it depends on the sample temperature and on the presence of defects and heterostructures.

\section{Effect of surface depletion and band bending}

\begin{figure}
	\includegraphics[width=\columnwidth]{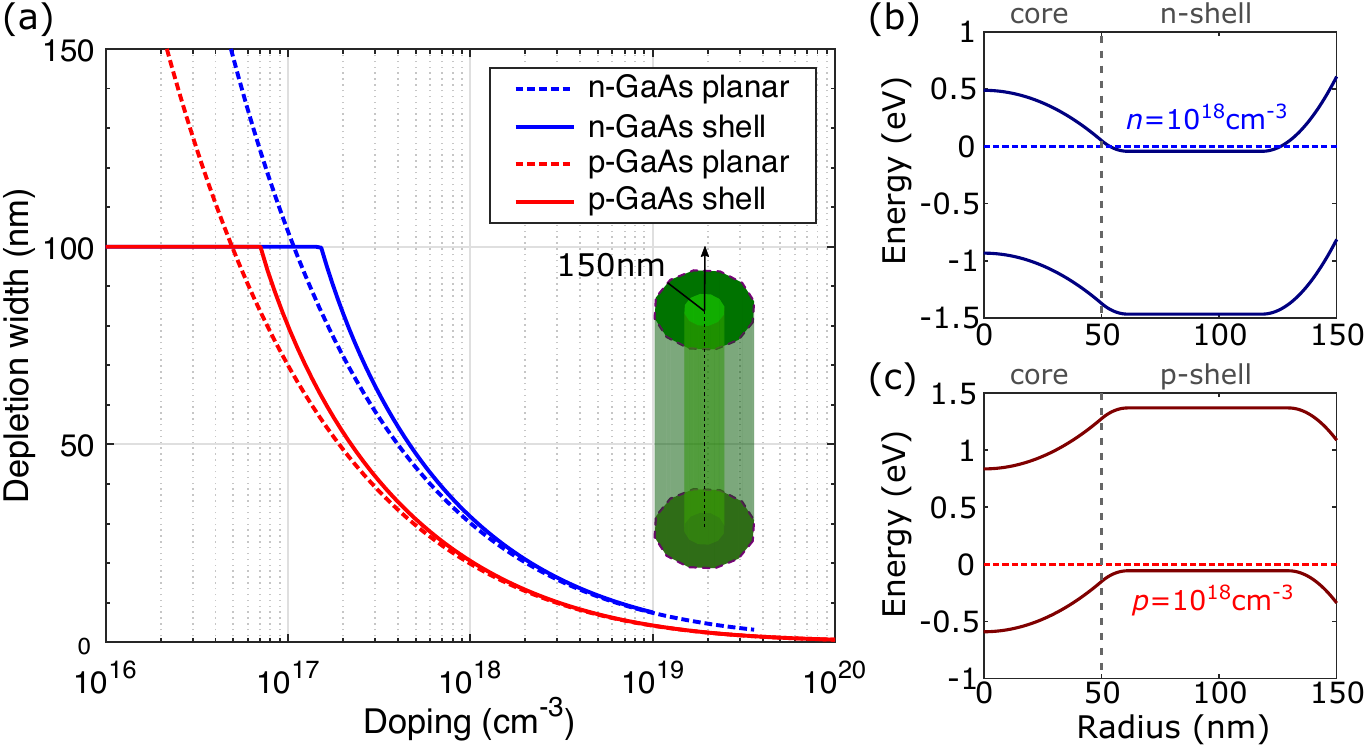}
	\caption{\label{fig:surface_depletion} (a) Calculation of the surface depletion width as a function of the doping concentration for p-GaAs (red) and n-GaAs (blue). Dashed lines correspond to bulk planar GaAs with a density of surface states $D_s=10^{13}$~cm$^{-2}$eV$^{-1}$ and a fixed charge neutrality level at 0.53~eV above the valence band maximum. Solid lines represent the shell width of the depletion region for GaAs NWs (total radius: 150~nm, shell thickness: 100~nm) with the same surface properties. (b,c) Band diagram along the NW radius for (b) n-doped and (c) p-doped GaAs NW.}
\end{figure}

For core-doped samples, an AlGaAs passivation shell was grown to increase the radiative efficiency of GaAs core, because thin GaAs NWs (diameter$<$130~nm) without passivation suffer from severe Fermi level pinning~\cite{Demichel:2010}. For shell-doped samples, a relatively thick shell allowed luminescence signal to be recorded without additional surface passivation. However, the effect of surface depletion and the junction formed with the undoped core may still influence the radiative process and are discussed in the following.

Figure~\ref{fig:surface_depletion}(a) shows the GaAs surface depletion width as a function of the doping concentration for thin films and for NWs modeled as infinite cylinders of 150~nm radius with uniform doping. It is calculated using the electrostatic model of Ref.~\cite{Chia:2012} with the charge neutrality level of GaAs taken to be 0.53~V above the valence band~\cite{Gozzo:1992}, and a density of surface states of 10$^{13}$~cm$^{-2}$eV$^{-1}$ corresponding to GaAs with native oxide~\cite{Aspnes:1983}. NWs behave as thin films at high doping levels, but present larger surface depletion width at low doping concentrations. The NWs are totally depleted below a certain concentration limit. Under e-beam excitation, the separation of quasi-Fermi levels decreases toward the surface due to surface recombination. Hence, the CL contribution from the surface depletion region should be significantly lower than from deeper regions.

In the case where an unintentionally doped core is present, the doped shell may be partly depleted near the core-shell interface. This is illustrated in Fig.~\ref{fig:surface_depletion}(b,c) for n-doped and p-doped shell. Free carriers can diffuse to the core, leading to band bending (bend up for n-shell and bend down for p-shell). Because of high doping levels in the shell and the small volume of the core, the effect of free carrier depletion in the shell should be negligible. Under e-beam excitation, most of the excess carriers are generated in the shell (depth of the generation volume $\sim$70~nm with a cut-off defined as 80\% of the maximum energy~\cite{Chen:2017}). In addition, carriers generated in the core would be separated due to the internal field, resulting in low CL contribution from the core. CL measurements on shell-doped samples should mainly probe the quasi-neutral region of doped shells.

\section{\label{sec:be_doped}{Be}-doped GaAs nanowires}

\begin{figure*}
	\includegraphics[width=\textwidth]{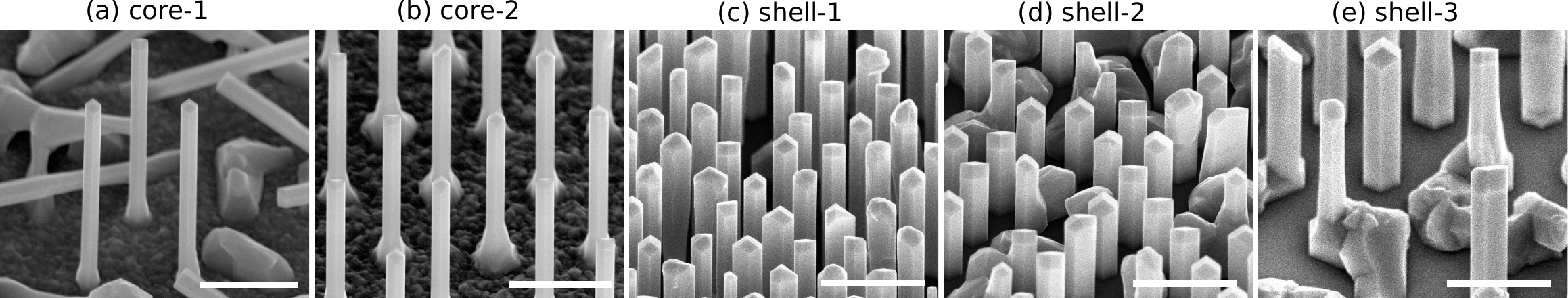}
	\caption{\label{fig:sem_be_nws}SEM images of GaAs:Be NWs on Si(111) substrates with a 45 degree sample tilt. (a,b) NWs with direct Be doping during the VLS growth. After crystallization of the Ga droplets, an AlGaAs shell was grown for surface passivation. (c,d,e) NW structures containing undoped GaAs core and Be-doped shell without additional surface passivation. Scale bars are 1~$\mu$m.}
\end{figure*}

\begin{figure*}
	\includegraphics[width=0.9\textwidth]{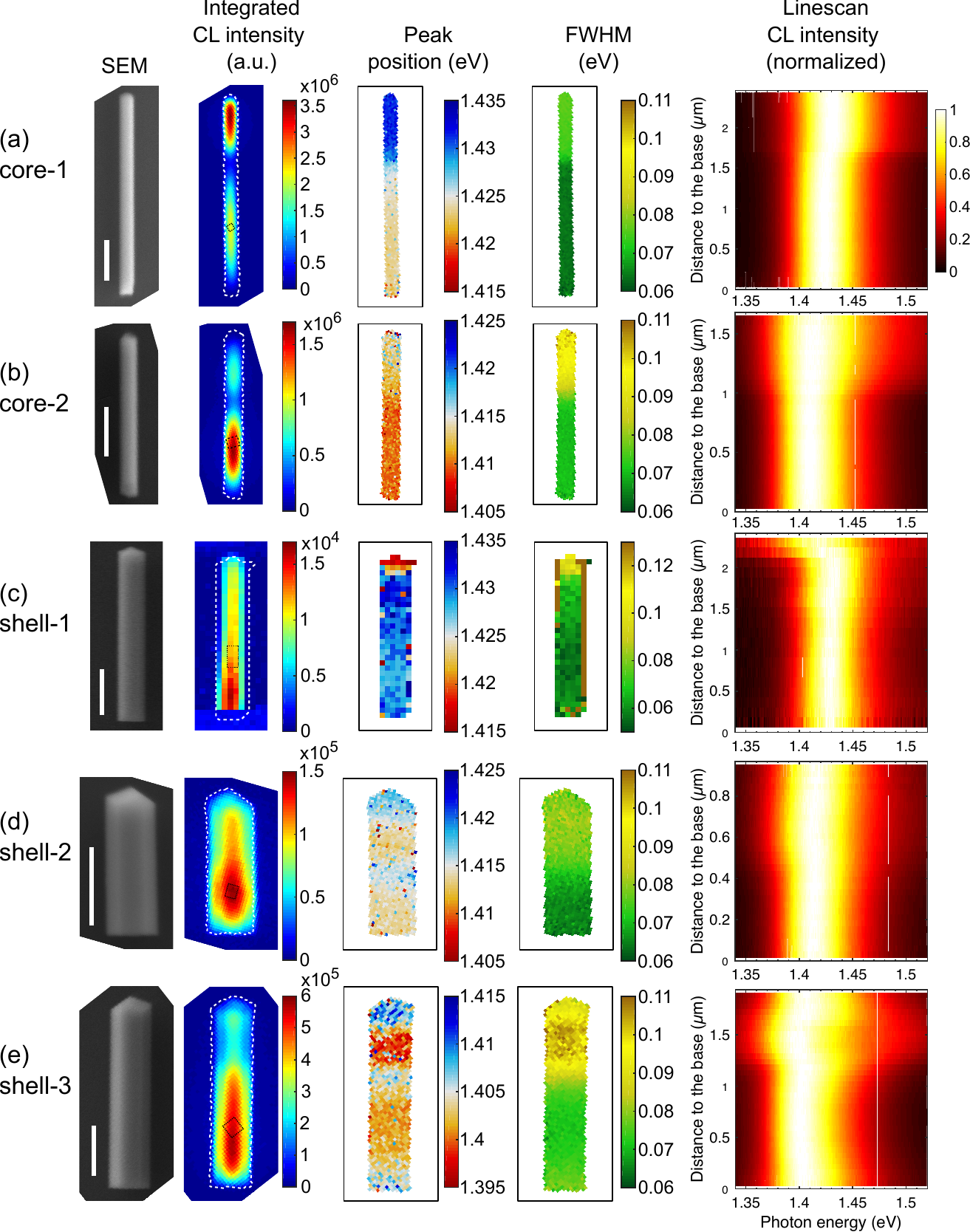}
	\caption{\label{fig:cl_map_be}Room-temperature CL maps of Be-doped GaAs NWs. For each CL measurement, SEM image of the NW and maps of the integrated CL intensity, the CL peak position and the FWHM are shown. One-dimensional linescans show the CL spectra along the nanowire axis in the right column. Most of the excess carriers are generated in the shell (depth $\sim$70~nm). The growth direction is from the bottom to the top of the NWs. Scale bars are 500~nm.}
\end{figure*}

The mechanism of Be incorporation in MBE-grown GaAs NWs by the VLS method is still unclear, due to the complexity introduced by the nanoscale liquid Ga droplets and the difficulty of doping characterization for single NWs. Early studies used electrical conductivity measurements along NWs and showed that Be impurities incorporate dominantly through the parasitic sidewall growth in the VS mode~\cite{Casadei:2013}, and this effect seems to dominate at low Be flux~\cite{Piton:2019}. Remarkably, Be incorporation through the three (112)A truncated facets at the liquid-solid interface was inferred using electron holography to evidence a three-fold symmetry of the electrical potential~\cite{Dastjerdi:2017}. On the other hand, homogeneous incorporation was observed using the reconstruction with atom probe tomography~\cite{Zhang:2018}. In our experiments, we observed rather homogeneous CL mapping for Be-doped NW core, suggesting no significant change of the incorporation along the nanowire axis in our growth conditions. For Be-doped NW shells, the situation should be similar to usual thin film growth and efficient doping incorporation was found using a high V/III ratio~\cite{Ojha:2016}.

Figure~\ref{fig:sem_be_nws} shows SEM images of as-grown NWs on Si substrates for the five Be-doped samples. Two of them are doped during the VLS growth (core-1 and -2), three are doped during the VS growth of NW shells (shell-1, -2 and -3). The statistics of the NW geometry are given in Table~\ref{tab:sample}. The growth parameters were kept identical except for the flux of Be dopant, but slight variations of the NW geometry and morphology are still observed. Be dopant is believed to have a surfactant effect which may cause irregular morphology and kinked NWs at high Be concentration~\cite{Zhang:2018}. In our samples, regular morphology was obtained along most of the NW length.

Figure~\ref{fig:cl_map_be} shows room-temperature (RT) CL measurements of Be-doped NWs. Every NW was measured with the same condition for electron excitation and luminescence collection, so their integrated CL intensities can be quantitatively compared. It allows for a direct comparison of the radiative efficiencies between samples. The color scales for the CL peak position and full-width at half maximum (FWHM) are adapted for each sample to visualize variations inside single NWs. For core-doped samples, two distinct regions with different CL characteristics are visible. The (top) \emph{droplet crystallization segment} of the NWs exhibits CL peaks with noticeable blueshift and broadening compared to the pure ZB segment, corresponding to the existence of a WZ segment. For shell-doped samples (shell-2 and shell-3), we can observe that the \emph{droplet crystallization segment} and the \emph{tip segment} show slightly different peak positions and broader FWHM as compared to the \emph{pure ZB segment}. These features may be caused by the crystal phase mixing and/or by doping inhomogeneities. In the following, we focus on the \emph{pure ZB segment}, seen as a homogeneous region in the middle of the NWs, in order to assess and compare the carrier density between samples.

\begin{figure}
	\includegraphics[width=\columnwidth]{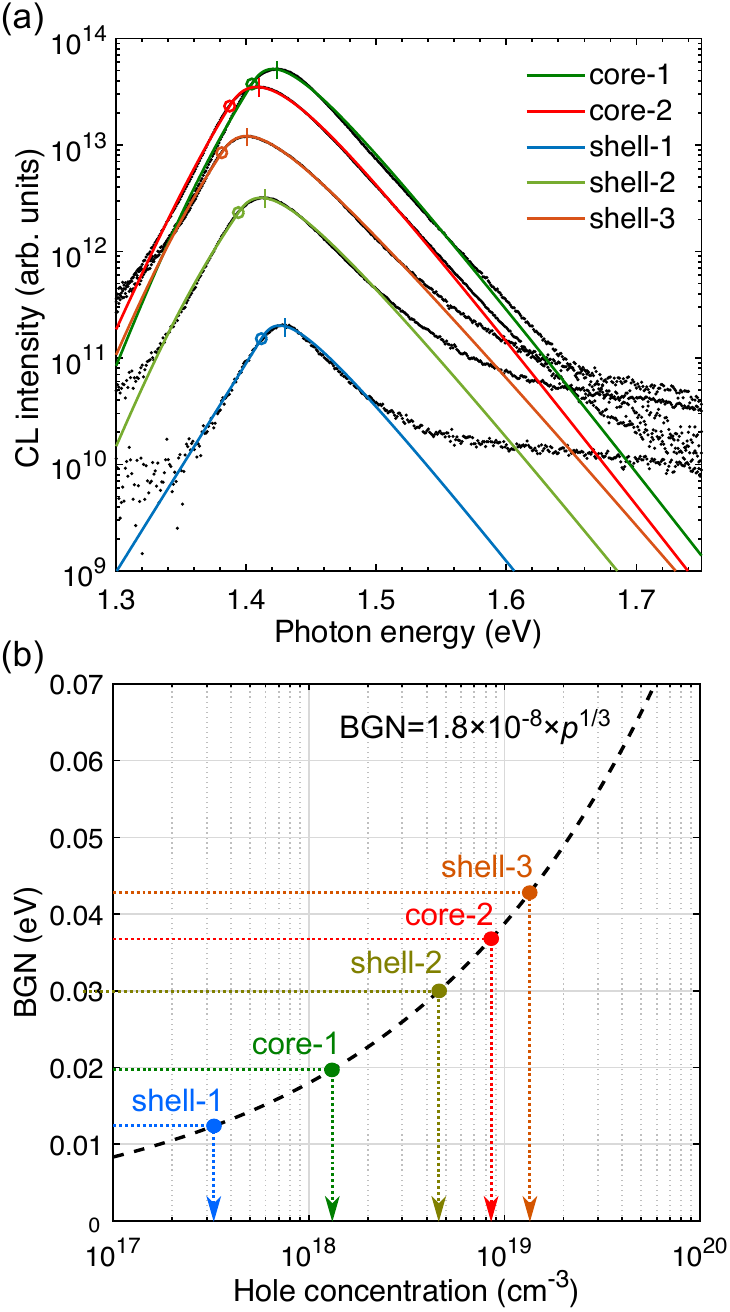}
	\caption{\label{fig:cl_log_be}(a) CL spectra of Be-doped GaAs NWs (black dots) and fits (colored lines). CL spectra are extracted from the middle homogeneous part of the CL maps (\emph{pure ZB segment}, rectangles in Fig.~\ref{fig:cl_map_be}). Vertical bars mark the CL peak positions, and open circles mark the bandgap deduced from the fit of the generalized Planck law. (b) Empirical relation used to access the hole concentration from the BGN values, from Ref.~\cite{Chen:2019a}.}
\end{figure}

\begin{table}
	\caption{CL characteristics of GaAs:Be NW samples. Peak position and FWHM are typical for CL spectra extracted from the middle homogeneous part of the NWs (\emph{pure ZB segment}). The bandgap $E_g$ and Urbach tail $\gamma$ are determined from the fit of CL spectra with the generalized Planck law. Experimental hole concentrations are obtained from the BGN values. For shell-1, the BGN value is too small to give a precise assessment of the hole concentration, the result is only indicative.}
	\label{tab:p_doping}
	\begin{ruledtabular}
	\begin{tabular}{lccccc}
		sample & peak & FWHM & $E_g$ & $\gamma$ & exp. hole conc.\\
		& eV & eV & eV & meV & cm$^{-3}$\\
		\hline
		core-1 & 1.423 & 0.065 & 1.404 & 13 & $1.3\times10^{18}$\\ 
		core-2 & 1.411 & 0.073 & 1.387 & 15 & $8.5\times10^{18}$\\ 
		shell-1 & 1.429 & 0.060 & 1.412 & 12 & ($3.3\times 10^{17}$) \\ 
		shell-2 & 1.413 & 0.070 & 1.394 & 15 & $4.6\times10^{18}$\\ 
		shell-3 & 1.401 & 0.073 & 1.381 & 15 & $1.3\times10^{19}$\\ 
	\end{tabular}
	\end{ruledtabular}
\end{table}

Average CL spectra are extracted from the homogeneous regions delimited by the small rectangles shown in Fig.~\ref{fig:cl_map_be} (integrated CL intensity maps). They are plotted in Fig.~\ref{fig:cl_log_be}(a) and can be compared quantitatively. The higher intensities are obtained with the core-doped samples, demonstrating the effective passivation effect of AlGaAs shells. The intensities for the shell-doped samples increase with the doping levels. In particular, the lowest doped NW (shell-1) exhibits a very low luminescence intensity due to a thinner shell and a larger surface depletion width. Redshift and broadening of CL spectra are consistent with an increased p-type doping, as expected from the Be flux (Table~\ref{tab:sample}).

CL spectra are fitted using the generalized Planck law and a parabolic absorption model convoluted with an Urbach tail, following the method described in Ref.~\cite{Chen:2019a}. Due to the small size of the NWs, we neglect the reabsorption effect so that the absorptivity is approximated by the absorption coefficient $\alpha(\hbar\omega)$~\cite{Greffet:2018}. The results are plotted in Fig.~\ref{fig:cl_log_be}(a) (colored lines). The fitted model provides the bandgap $E_g$ (open circles in Fig.~\ref{fig:cl_log_be}(a)) and energy width of the Urbach tail $\gamma$. Both values depend mainly on the low-energy part of the luminescence spectra. Table~\ref{tab:p_doping} lists the Be-doped GaAs NW samples, the characteristics of CL spectra (peak energy and FWHM), $E_g$ and $\gamma$. Experimental hole concentrations are deduced from the bandgap narrowing (BGN) values: $\text{BGN}=1.424-E_g(\text{eV})$, and the empirical relation determined in reference~\cite{Chen:2019a}, see Fig.~\ref{fig:cl_log_be}(b). For shell-1, the BGN value is too small to give a precise assessment of the hole concentration, which is probably lower than $1\times 10^{18}$~cm$^{-3}$. In this range, measurements may be also impacted by surface depletion due to the absence of AlGaAs passivation shell (see Fig.~\ref{fig:surface_depletion}), so the result given for shell-1 in Table~\ref{tab:p_doping} and Fig.~\ref{fig:cl_log_be}(b) is only indicative.

The carrier concentrations determined by CL (Table~\ref{tab:p_doping}) are in very good agreement with the expected doping levels (Table~\ref{tab:sample}). This is particularly the case for core-doped samples, suggesting that Be atoms incorporate through the VLS process in our growth conditions, and that Be has an activation efficiency close to 100~\%. The alternative mechanism of Be incorporation during the intentional axial growth is the direct incorporation from the vapor into the NW sidewalls (VS mode)~\cite{Casadei:2013,Piton:2019}, corresponding to a parasitic radial growth. This mechanism should lead to a gradual decrease of dopant concentration from the bottom to the top of the pure ZB segment. Since we do not observe such a concentration gradient, the VS process must be negligible in the core-doped samples. For shell-doped samples, the small discrepancies between the theoretical and experimental doping concentrations may be due to the uncertainty of the growth rate, complex orientation and shadowing of the impinging dopant flux, lower activation, and surface depletion for low concentration. Overall, Be-doping in the core or shell of GaAs NWs can be accurately controlled through CL calibration.

\begin{figure}
	\includegraphics[width=\columnwidth]{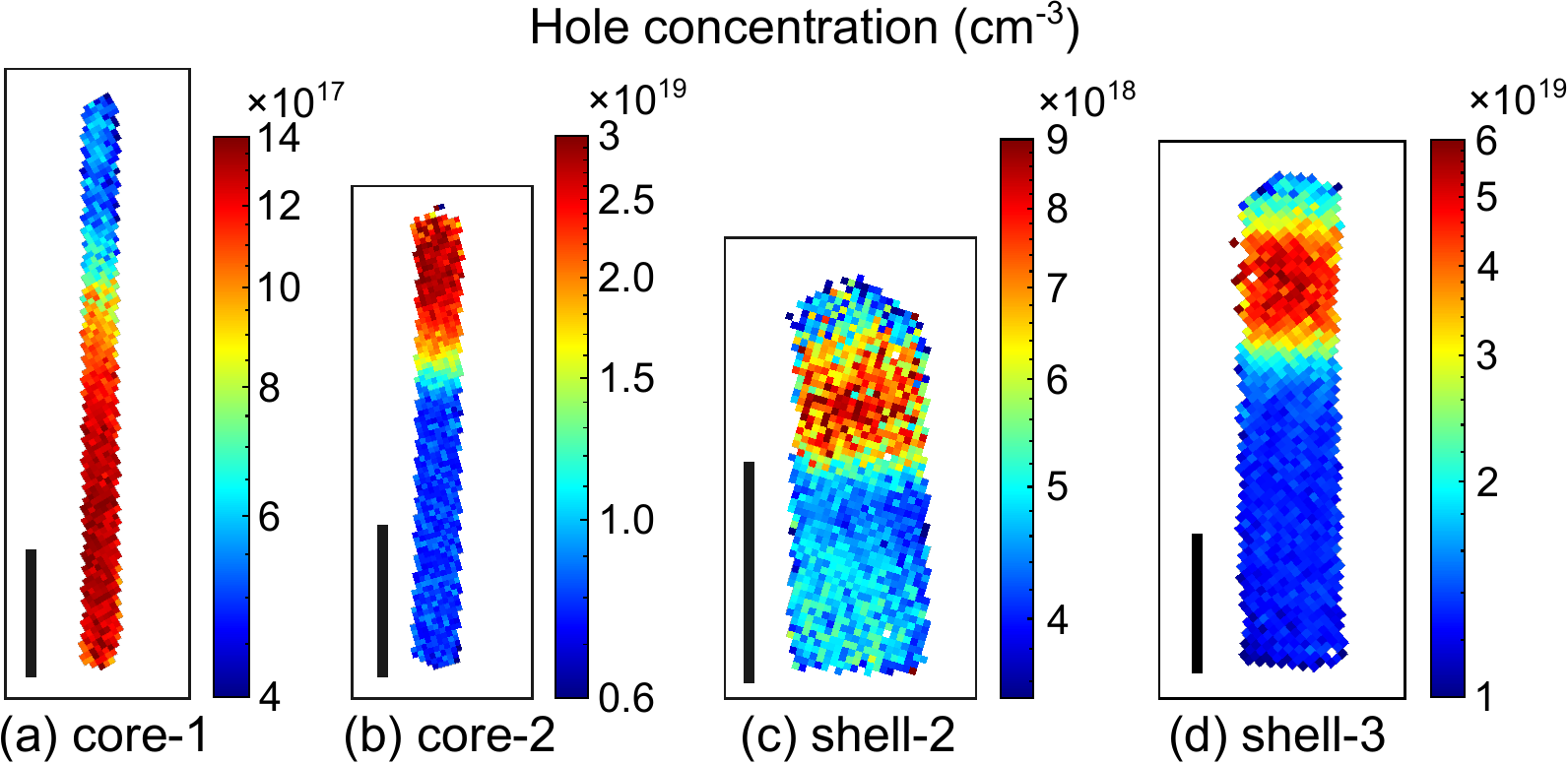}
	\caption{\label{fig:dopingmap_be} Maps of hole concentration, determined with a pixel-by-pixel analysis of the CL spectra. The growth direction is from the bottom to the top of the NWs. Scale bars are 500~nm.}
\end{figure}

Figure~\ref{fig:dopingmap_be} shows high-resolution maps of the hole concentration determined with a pixel-by-pixel analysis, following the previous method shown in Fig.~\ref{fig:cl_log_be}: each spectrum is fitted with the generalized Planck law, and the hole concentration is deduced from the resulting bandgap. The sample shell-1 was not included because of the lack of accuracy of the method for low BGN. For the other samples, the resulting maps exhibit homogeneous hole concentration in the pure ZB segments of the nanowires (red regions in Fig.~\ref{fig:dopingmap_be}), and the values are in very good agreement with the previous analysis and Table~\ref{tab:p_doping}.

Closer to the tip, the region correspond to mixed phases and a WZ segment. However, the relation between the BGN and the hole concentration was determined on ZB GaAs reference layers~\cite{Chen:2019a}.
In the region of crystal phase mixing and in the WZ segment, the CL characteristics are also affected by these structural switches. The variation of doping level shown in Fig.~\ref{fig:cl_map_be} must be corrected from these effects and the values outside the ZB segment must be considered with caution. Further analysis will be required to extend the determination of carrier concentration to the GaAs WZ.

\section{\label{sec:si_doped}{Si}-doped GaAs nanowires}

The incorporation of Si dopants is a contentious topic in GaAs NW growth due to its amphoteric behavior. It is normally utilized as an n-type dopant for GaAs thin films with (001) surface orientation, and typical free electron concentration up to $7\times 10^{18}$cm$^{-3}$ without dopant compensation can be achieved in MBE systems~\cite{Neave:1983}.

However, the incorporation mechanism differs largely in the presence of a liquid phase. Si-doped GaAs grown by liquid phase epitaxy (LPE) is mainly compensated~\cite{Hurle:1999} and {p}-type doping was observed in GaAs NWs grown by VLS~\cite{Dufouleur:2010,Ketterer:2010}. On the other hand, Si doping on NW \{110\} facets is believed to produce n-type conductivity~\cite{Dimakis:2012}, and Si incorporation on (110) GaAs surface has been shown to depend strongly on the As/Ga flux ratio. Low As/Ga ratio or high growth temperature may result in highly compensated n-type or even p-type conductivity~\cite{Tok:1998}. Here, we have grown Si-doped GaAs NW shells at low temperature ($T<500^\circ$C) to investigate the n-doping characteristics by CL mapping.

\begin{figure}
	\includegraphics[width=\columnwidth]{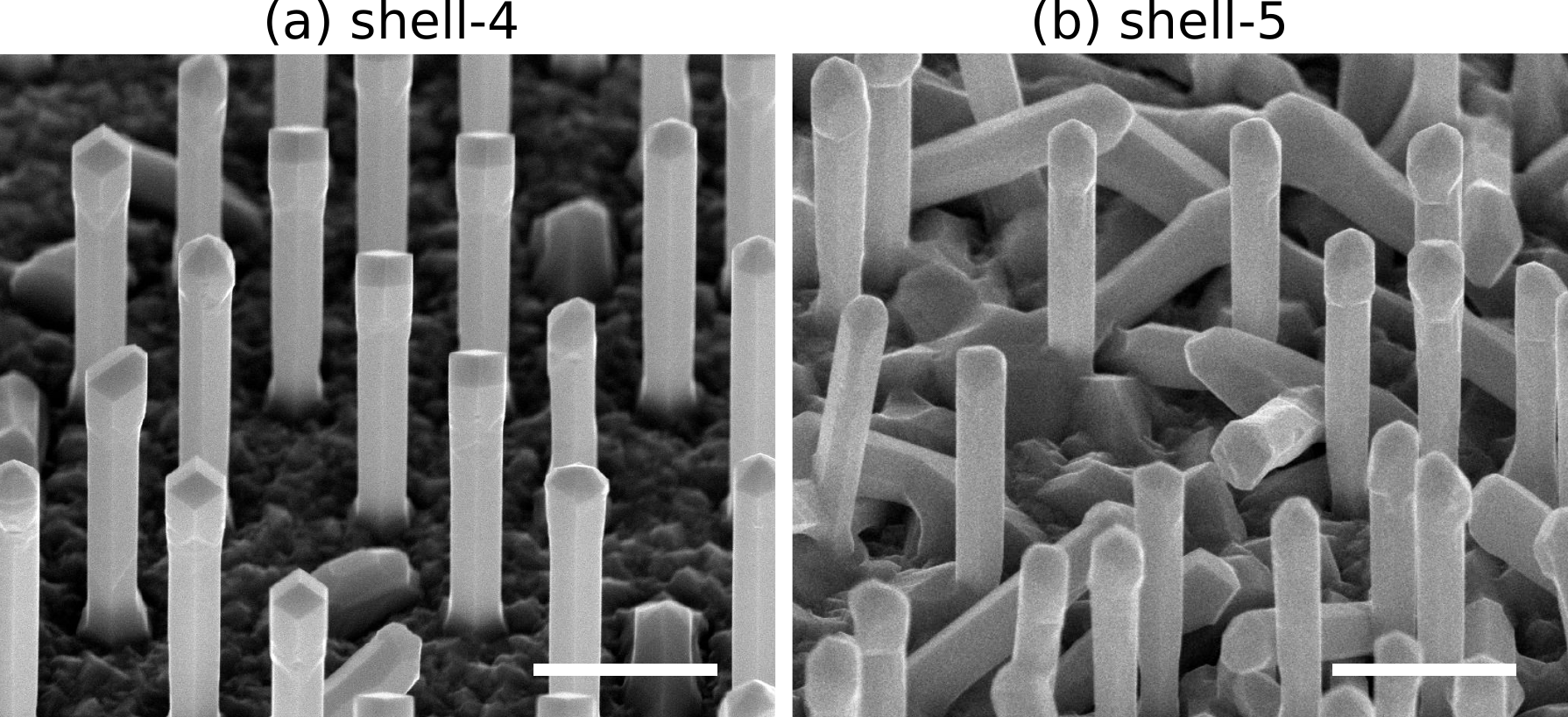}
	\caption{\label{fig:sem_si_nws} SEM images of GaAs:Si NWs on Si(111) substrates with a 45 degree sample tilt. GaAs NWs are composed of undoped core and Si-doped shell without additional surface passivation. Scale bars are 1~$\mu$m.}
\end{figure}

\begin{figure*}
	\includegraphics[width=0.8\textwidth]{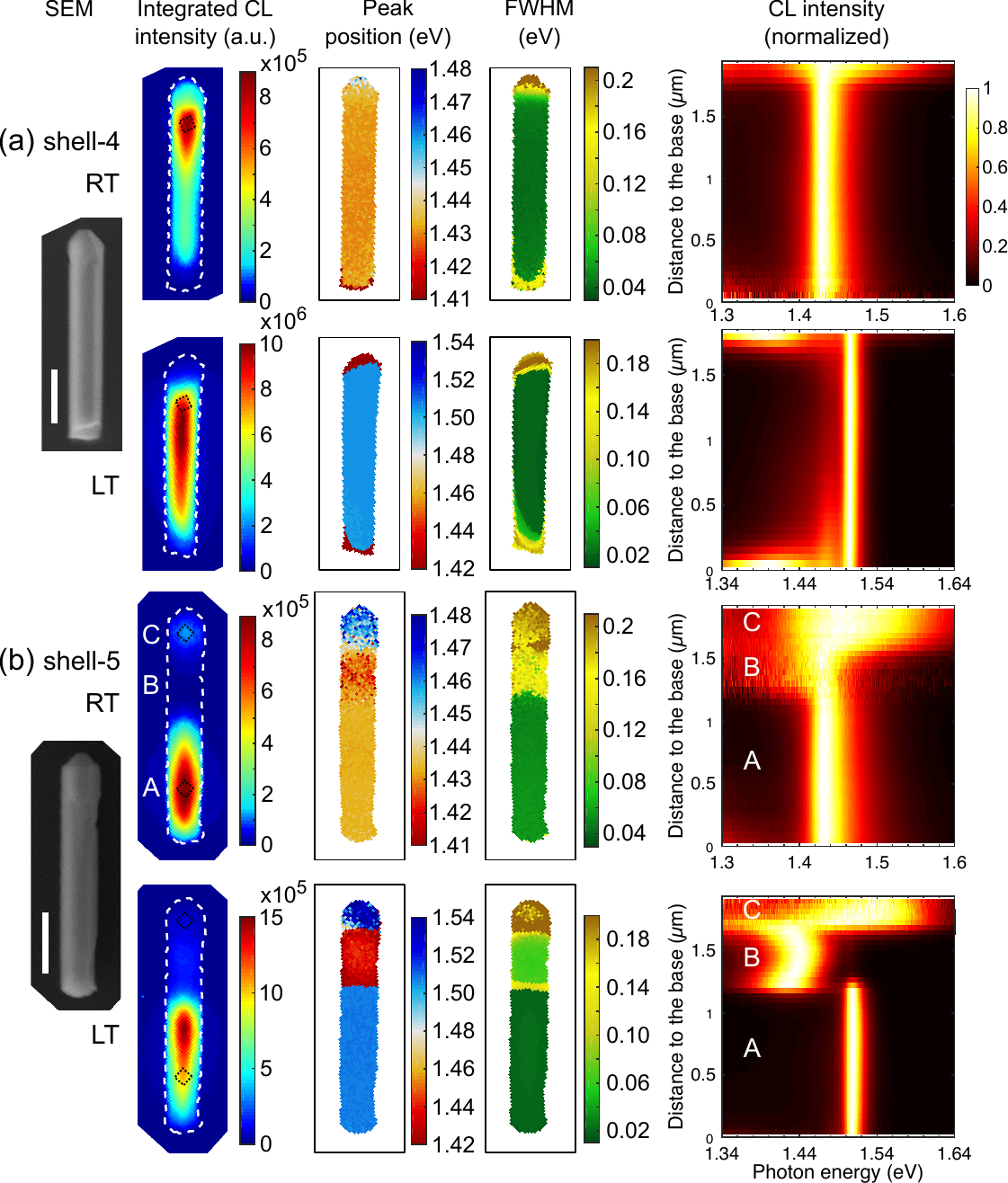}
	\caption{\label{fig:cl_map_si}SEM images, RT and LT CL maps of Si-doped GaAs NWs. For each CL measurement, maps of the integrated CL intensity, the CL peak position and the FWHM are shown. One-dimensional linescans show the CL spectra along the nanowire axis in the right column. Most of the excess carriers are generated in the shell (depth $\sim$70~nm). The growth direction is from the bottom to the top of the NWs. Scale bars are 500~nm.}
\end{figure*}

Figure~\ref{fig:sem_si_nws} shows SEM images of two samples (shell-4 and shell-5). Their growth parameters, geometry and expected [Si] concentrations in the shell are given in Table~\ref{tab:sample}. The presence of Si atoms during the shell growth or/and the reduced growth temperature also causes parasitic growth of GaAs crystals on the \ce{SiO2} mask. The morphology of GaAs:Si NWs is regular along most of the wire length, except at the top where twin planes, stacking faults and mixed crystal phases may be induced by the crystallization of the Ga droplets \cite{Dastjerdi:2016,Himwas:2017}. For shell-4, the consumption of liquid Ga droplets was conducted at higher temperature (610$^\circ$C) and lower As flux (roughly one sixth of the value used during the VLS growth) to suppress the defect formation at the tip~\cite{Dastjerdi:2016}.

Figure~\ref{fig:cl_map_si} shows SEM images, RT and low-temperature (LT, 20~K) CL maps of Si-doped GaAs NWs. Shell-4 presents homogeneous CL characteristics (peak position and FWHM) along the entire wire. On the other hand, CL maps of shell-5 reveal a region (labeled B) in the top half where the CL intensity is reduced and the CL peak is redshifted. It corresponds probably to the core region grown during the droplet crystallization (\emph{droplet crystallization segment}). The top part (labeled C) exhibits wide and blueshifted CL features. This part (\emph{tip segment}) was likely formed on top of the droplet crystallization segment, during the shell growth. It is worth noting that its emission spectra shift to higher energies (beyond 1.54 eV at LT) and are significantly broader than in the WZ/ZB mixed phase regions. The rest of the wire (bottom to middle part, labeled A) remains homogeneous (\emph{pure ZB segment}).

\begin{figure}
	\includegraphics[width=\columnwidth]{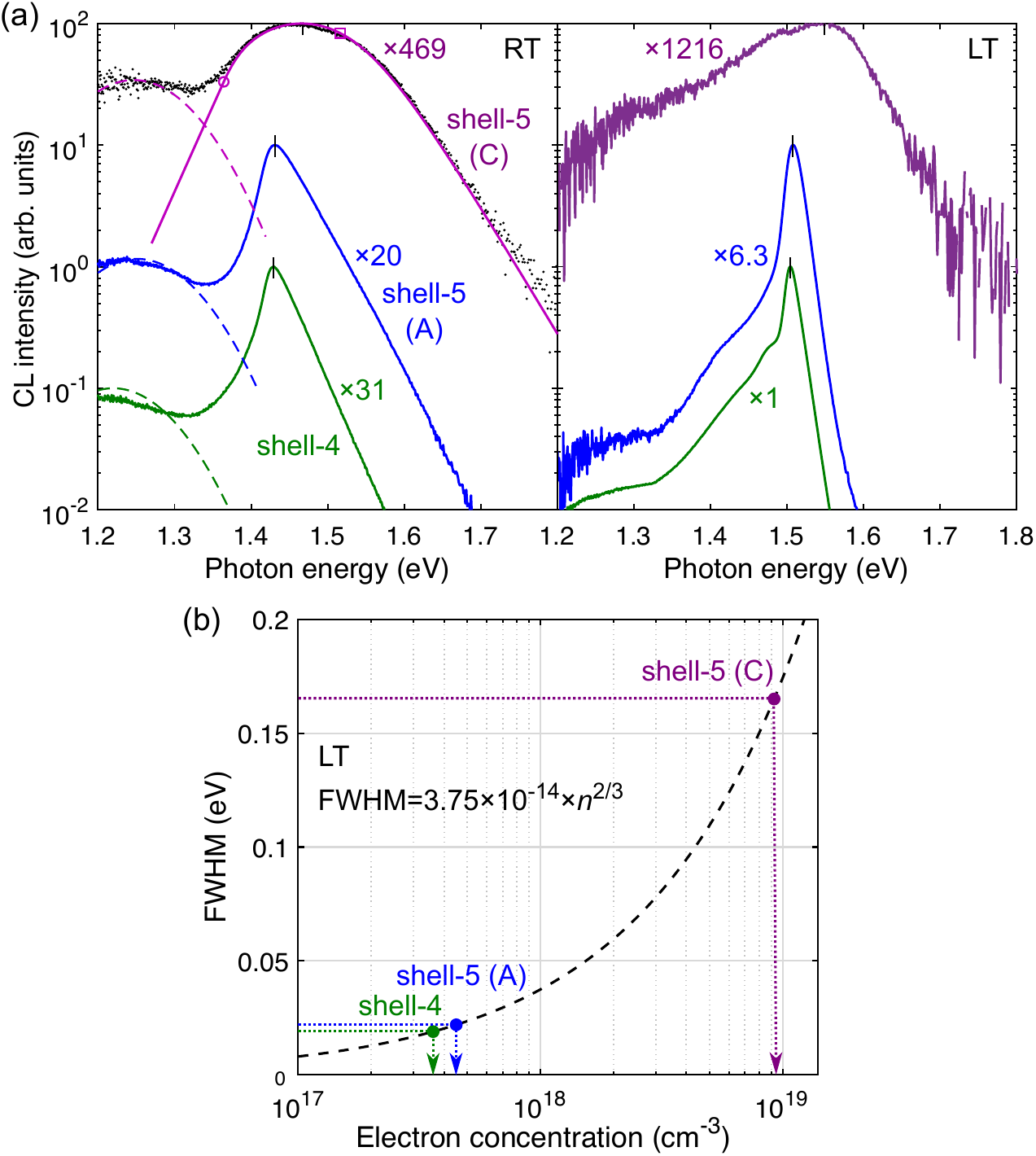}
	\caption{\label{fig:cl_log_si}(a) Room-temperature (RT, left) and low-temperature (LT, right) CL spectra of Si-doped GaAs NWs. CL spectra of shell-5 are extracted from the bottom and the top of the NW (regions A and C, respectively, rectangles in Fig.~\ref{fig:cl_map_si}). The RT CL spectrum of shell-5 top (C) is fitted (solid curves) to extract the bandgap (open circle) and the electron Fermi level (open square). (b) Empirical relation used to access the electron concentration from LT FWHM of CL spectra, from Ref.~\cite{Chen:2019a}.}
\end{figure}

\begin{table}
	\caption{\label{tab:n_doping}RT and LT CL characteristics of GaAs:Si NW samples. Electron Fermi levels cannot be extracted reliably from the luminescence lineshape at low concentrations. Experimental electron concentrations are assessed from the LT FWHM.}
	\begin{ruledtabular}
		\begin{tabular}{lccccc}
			& \multicolumn{2}{c}{RT} & \multicolumn{2}{c}{LT} & \\
			\cline{2-3}\cline{4-5}
			sample & peak & FWHM & peak & FWHM & exp. electron conc. \\
			& eV & meV & eV & meV & cm$^{-3}$ \\
			\hline
			shell-4 & 1.430 & 42 & 1.505 & 19 & $3.6\times10^{17}$ \\ 
			shell-5 (A) & 1.431 & 54 & 1.508 & 22 & $4.5\times10^{17}$ \\ 
			shell-5 (C) & 1.468 & 191 & 1.549 & 165 & $9.2\times10^{18}$ \\ 
		\end{tabular}
	\end{ruledtabular}
\end{table}

In Fig.~\ref{fig:cl_log_si}(a), we compare CL spectra extracted from shell-4 and from the bottom homogeneous region (A) and the top part (C) of shell-5. They are normalized and shifted vertically for clarity. The low-energy peak observed on RT CL spectra around 1.2--1.3~eV (dashed lines in Fig.~\ref{fig:cl_log_si}(a) left) is attributed to Si-related complex defects~\cite{Kressel:1969,Chiang:1975,Pavesi:1992}. The main emission peak can be fitted following the method developed in Ref.~\cite{Chen:2019a}. However, for the two thinner luminescence peaks of shell-4 and shell-5 (A), the electron concentration is probably below the degeneration threshold and the electron Fermi level $E_{fc}$ cannot be deduced accurately. In contrast, the top region (C) of shell-5 exhibits a broad, blueshifted CL spectrum, which is consistent with a high electron concentration in this region. The crystal phase mixing and defect luminescence cannot explain such a large blueshift of the CL peak (peak position 40 meV higher than the RT undoped GaAs bandgap). The main emission peak (RT) is nicely fitted with $E_g=1.364$~eV, $E_{fc}$=0.152~eV, $\gamma=29$~meV, and $kT=38$~meV. An effective higher carrier temperature is reflected from the slower decrease of the high-energy tail of the CL spectrum, as compared with the two other CL spectra. This may be related to the k-conservation in the radiative recombination of electrons above the degenerate Fermi level with holes that are not thermalized to the band edge~\cite{Mergenthaler:2017}. The electron Fermi level of 152~meV corresponds to an electron concentration of 5.1$\times 10^{18}$~cm$^{-3}$~\cite{Chen:2019a}.

To provide a more accurate assessment of the carrier density, CL measurements of the same NWs have been performed at LT. Their spectra are shown in Fig.~\ref{fig:cl_log_si}(a) (right), and the luminescence characteristics (peak position and FWHM) are summarized in Table~\ref{tab:n_doping}. The electron concentration is estimated using the LT FWHM and the empirical relation established in Ref.~\cite{Chen:2019a}, see Fig.~\ref{fig:cl_log_si}(b). We find an electron concentration around $4\times 10^{17}$~cm$^{-3}$ for shell-4 and shell-5 (\emph{pure ZB segment}, region A). The FWHM maps in Fig.~\ref{fig:cl_map_si} demonstrate the excellent homogeneity of the doping in the pure ZB segment. For the \emph{tip segment} of shell-5 (region C), we find an electron concentration of $9.2\times 10^{18}$~cm$^{-3}$. We note that this value may be slightly overestimated due to low-energy luminescence from the defects. Comparing with the fit of the RT CL spectrum, we infer a high carrier density of $n=5\text{--}9\times 10^{18}$~cm$^{-3}$ at the top of shell-5.

We notice a discrepancy of nearly one order of magnitude between the target doping levels (Table~\ref{tab:sample}) and the values deduced from CL  measurements (Table~\ref{tab:n_doping}). Lowered doping efficiency in the NW shell was also observed in Te-doped GaAs NWs~\cite{Goktas:2018}. Surface depletion~\cite{Dimakis:2012} or Si dopant compensation due to (110) surface orientation~\cite{Tok:1998} may be responsible for the observed lower carrier density determined by CL. Growth issues such as inhomogeneous distribution of molecular beam or shadowing effect~\cite{Oehler:2018} may also hinder the incorporation of Si dopants.
Indeed the NW sidewalls are vertical. Since the substrate rotates during shell growth (2.5 s period), these facets are not constantly and simultaneously exposed to Ga, As and Si fluxes. In particular, with the effusion cell configuration of our MBE chamber, Si and Ga fluxes impinge a vertical facet when the As flux on this facet is shadowed. This situation is favorable to the incorporation of a part of Si impurities as acceptors~\cite{Tok:1998} and may result in a high degree of compensation. This is not the case at the NW tip, which receives the three fluxes simultaneously at any angle of sample rotation. This may explain the much higher carrier density measured in the tip segment of shell-5. Further understanding of microscopic crystal growth mechanisms in various configurations will be required to improve the NW growth conditions for device applications.

\section{Conclusion}

In this paper, we have investigated Be-doped and Si-doped GaAs NWs grown by MBE in different conditions, and with different target doping concentrations. CL mapping allows to identify defective regions in single NWs and to extract spectra from homogeneous regions for further analysis. For Be-doped GaAs NWs, CL spectra progressively broaden and redshift with increasing Be flux. The bandgap narrowing is determined by fitting RT spectra with the generalized Planck law, and used to assess quantitatively the hole concentration. It is found to be in very good agreement with expected doping levels, and a pixel-by-pixel analysis provides high-resolution maps of the hole concentration and demonstrates the excellent homogeneity of the doping.
For Si-doped GaAs NWs, increased doping concentration induces a blueshift and spectral broadening related to the well-known Burstein-Moss effect. In the homogeneous region in the middle of the NWs, we infer electron concentration of $3\times 10^{17}$ to $6\times 10^{17}$~cm$^{-3}$ from LT FWHM of CL spectra. Luminescence from a defect band at lower energies indicates Si dopant compensation, which may be induced by the inhomogeneous fluxes on the vertical facets of NWs during the growth. We also unveil signatures of higher electron concentrations (5--9$\times 10^{18}$~cm$^{-3}$) at the tip of NWs, providing a route to overcome the doping limitation of previous Si-doped GaAs NWs.

In conclusion, we have shown that high-resolution CL mapping and spectral analysis provide useful information for contactless assessment of carrier densities at the nanoscale. This technique can be extended to other semiconductor and applied to numerous  nanostructures and devices. In addition, in a single high-resolution CL mapping experiment, one can obtain other insights about surface passivation, carrier lifetime and diffusion length, and the composition of a wide variety of ternary or quaternary compound semiconductors.

\section*{Acknowledgments}
The authors warmly thank Gilles Patriarche for TEM experiments. This project has been supported by the French government in the frame of the "Programme d'Investissement d'Avenir" - ANR-IEED-002-01 and by the ANR projects Nanocell (ANR-15-CE05-0026) and Hetonan (ANR-15-CE05-0009).
This work was also partly funded in the frame of the EMPIR 19ENG05 NanoWires project. The EMPIR (European Metrology Programme for Innovation and Research) initiative is co-funded by the European Union's Horizon 2020 research and innovation programme and the EMPIR Participating States.

\appendix*
\section{}

\begin{figure*}
	\includegraphics[width=\textwidth]{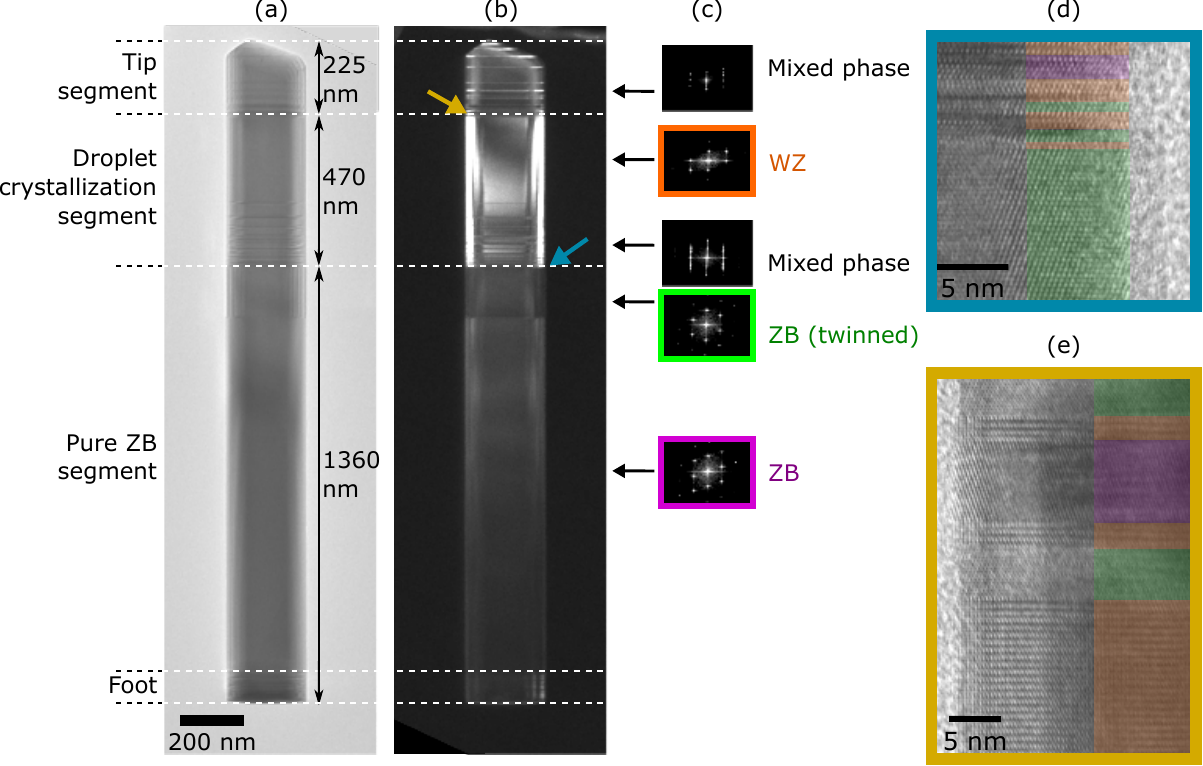}
	\caption{\label{fig:TEM}(a) TEM-BF micrograph of a NW with a similar core growth procedure as the samples in the article, viewed along the 〈110〉 zone-axis (b) TEM-DF micrograph highlighting the WZ phase. Above the NW foot, 3 distinctive segments can be identified from the crystal phase. (c) The corresponding Fourier transforms (FFT) are shown to the right. (d,e) High resolution TEM-BF images at the locations indicated by colored arrows, showing the cubic zinc-blende (ZB) twinned crystals (green and pink color) and the hexagonal wurtzite (WZ) phase (orange color).}
\end{figure*}

During the growth of self-catalyzed GaAs nanowires (NWs), the consumption of Ga droplets may create NW segments with different crystal structures. Here, we present NWs grown using the same conditions as samples shell-1, -2, and -3 of the main text for the core and for the catalyst consumption, and we analyze their structure by transmission electron microscopy (TEM). Figure~\ref{fig:TEM} shows the bright-field TEM (TEM-BF) and dark-field TEM (TEM-DF) images of a typical NW dispersed on a TEM grid. Four segments can be identified: (1) the NW foot with a twinned cubic (ZB) structure, (2) the main NW segment with a near twin-free ZB crystal structure, (3) a mixed phase segment with both cubic, twinned cubic and an extended hexagonal (WZ) section, (4) the final NW tip with cubic and twinned cubic phases.

The presence of the hexagonal wurtzite (WZ) phase in segment (3) is consistent with the shrinking of the Ga droplet exposed to As4 flux during the droplet consumption step, thus reducing its contact angle. It is widely reported that the cubic zinc-blende (ZB) phase is stable for large  and low contact angles, while a narrow interval near $90^{\circ}$ favors the nucleation of the WZ phase~\cite{Jacobsson2016}. Similar extended WZ segment of sandwiched between twinned ZB sections have been reported during the catalyst crystallization of self-catalyzed GaAs NWs in the literature~\cite{Priante2013,Kim2012a}.

\bibliography{CL_doping_nanowire}

\end{document}